\documentclass[twocolumn,secnumarabic,amssymb, amsmath, nofootinbib,tightenlines,
nobibnotes, aps, prl,epsfig]{revtex4}
\usepackage{graphicx}
\usepackage{dcolumn}
\usepackage{bm}
\begin{document}
\preprint{APS/123-QED}
\title{An approximate approach to the nonlinear DGLAP evaluation equation }

\author{G.R.Boroun}%
 \email{ grboroun@gmail.com; boroun@razi.ac.ir }
 \author{S.Zarrin}%
\affiliation{ Physics Department, Razi University, Kermanshah
67149, Iran}
\date{\today}
\begin{abstract}
We determined the effects  of the first nonlinear corrections to
the gluon distribution using the solution of the QCD nonlinear
Dokshitzer-Gribov-Lipatov-Altarelli-parisi (NLDGLAP) evolution
equation at small $x$. By using a Laplace-transform technique, the
 behavior of the gluon distribution is obtained by
solving the Gribov, Levin, Ryskin, Mueller and Qiu (GLR-MQ)
evolution equation with the nonlinear shadowing term incorporated.
We show that the strong rise that is corresponding to the linear
QCD evolution equations at small-$x$ can be tamed by screening
effects. Consequently, the nonlinear effects for the  gluon
distributions are calculated and compared with the results for the
integrated gluon density from the Balitsky- Kovchegov(BK)
equation. The resulting analytic expression allow us to predict
the shadowing correction to the logarithmic derivative
$F_{2}(x,Q^{2})$ with respect to ${\ln}Q^{2}$ and to
compare the results with H1 data and a QCD analysis fit.\\
\end{abstract}
 \pacs{***}
\keywords{****} 
\maketitle
\subsection{1.Introduction}

Perturbative QCD (PQCD) predicts a strong rise of the gluon
density $xg(x,Q^{2})$ at a small value of the Bjorken variable
$x$(${<<} 1$), where ,as usual, $x$ denotes the momentum fraction
carried by the gluon and $Q^{2}$ is the scale at which the
distribution is probed. This fact that gluon density increase as
$x$ decreases follows directly from linear evolution equations
[1-5]. This strong rise can eventually violate unitarity and so it
has to be tamed by screening effects. This behavior necessitates a
chance in the evolution equation in the region of small-$x$. At
small-$x$, the problem is more complicated since recombination
processes between gluons in a dense system have to be taken into
account, and so it has to be tamed by screening effects. These
screening effects are provided by multiple gluon interaction which
leads to the nonlinear terms [6-7] in the Dokshitzer- Gribov-
Lipatov- Altarelli- Parisi (DGLAP) evolution equations [1-3].
These nonlinear terms reduce the growth of the gluon distribution
in this kinematic region where $\alpha_{s}$ is still small but the
density of partons becomes so large. Gribov, Levin, Ryskin,
Mueller and Qiu (GLR-MQ)[6,8] performed a detailed study of this
region. They argued that the physical processes of interaction and
recombination of partons become important in the parton cascade at
a large value of the parton density, and that these shadowing
corrections could be
expressed in a new evolution equation. This is the so-called GLR-MQ equation.\\
 The main characteristic of this equation is
that it predicts a saturation of the gluon distribution at very
small-$x$ [8]. This equation
was based on two processes in a parton cascade:\\
 i)The emission
induced by the QCD vertex $G{\rightarrow}G+G$ with the probability
which is proportional to $\alpha_{s}\rho$ where
$\rho(=\frac{xg(x,Q^{2})}{{\pi}R^{2}})$ is the density of gluon in
the transverse plane, ${\pi}R^{2}$ is the target area, and $R$ is the size of the target which the gluons populate;\\
ii)The annihilation of a gluon by the same vertex
$G+G{\rightarrow}G$ with probability which is proportional to
$\frac{\alpha_{s}^{2}\rho^{2}}{Q^{2}}$, where $\alpha_{s}$ is
the probability of the processes.\\

When $x$ is not too small, only the emission of new partons is
essential because $\rho$ is small and this emission is described
by the DGLAP evolution equations. Since gluon recombination
introduces a negative correction to the DGLAP evolution equations,
 the signal of its presence is a decrease of the scaling
violations as compared to the expectations obtained from the DGLAP
equations. However as $x\rightarrow0$ the value of $\rho$ becomes
so large that the annihilation of partons becomes important. This
picture allows us to write the GLR-MQ equation for the gluon
distribution function behavior at small-$x$ [10-11]. The shadowing
corrections arise from fusion of two gluon ladders, and they
modify the evolution equations of gluons as
\begin{eqnarray}
\frac{{\partial}xg(x,Q^{2})}{{\partial}{\ln}Q^{2}}&=&\frac{{\partial}xg(x,Q^{2})}{{\partial}{\ln}Q^{2}}|_{DGLAP}\nonumber\\
&&-\frac{\alpha^{2}_{s}\gamma}{R^{2}Q^{2}}\int^{1}_{\chi}\frac{dy}{y}[yg(y,Q^{2})]^{2},
\end{eqnarray}
where the first term is the standard DGLAP result[1-3], linear in
the parton distribution functions (PDFs) and the second term is
the 2-gluon density. Thus the equation (1) becomes nonlinear in
$xg$. In this equation, $\gamma$ is equal to $\frac{81}{16}$ for
$N_{c}=3$. Here the representation for the gluon distribution
$G(x)=xg(x)$ is used, where $g(x)$ is the gluon density and
$\chi=\frac{x}{x_{0}}$, where $x_{0}(=0.01)$ is the boundary
condition that the gluon distribution joints smoothly onto the
unshadowed region.  We neglect the quark-gluon emission diagrams
due to their little importance in the gluon rich small-$x$ region
and we work under an approximation of neglecting contribution
 from the high twist gluon distribution $G_{HT}(x,Q^{2})$ [6].\\
 In
the present paper, we extend the method using a Laplace-transform
technique as proposed at Refs.[12-14]. Here we derive solution for
the gluon distribution function in terms of the initial condition
using the Laplace transform method, then extended this method for
obtain an analytical solution of the GLR-MQ equation as the
shadowing corrections to the gluon distribution functions can be
determine in terms of the initial gluon distribution at
$Q_{0}^{2}$.\\
 We perform the check of our calculations by comparing the determined
gluon density function with the solution of the more precise and
more complicated  BK equation [15-17]. Here the BK equation was
derived for the deep inelastic scattering of the virtual photon on
a large nucleus by the resummation of multiple Pomeron exchanges
in the leading logarithmic approximation and in the large $N_{c}$
limit. It is an equation for the imaginary part of the forward
scattering amplitude of the $q\overline{q}$ dipole on the nucleus.
In the infinite momentum frame, the BK equation resumes fan
diagrams with the BFKL [18-20] ladders, corresponding to the
Pomeron, splitting into two ladders. This type of summation was
originally proposed by GLR in the double logarithmic approximation
(DLA) ($\alpha_{s}<<1$ and
$\alpha_{s}ln(\frac{1}{x})ln(\frac{Q^{2}}{\Lambda^{2}}){\sim}1$).
The non-linear evolution equation for the amplitude reduce to the
GLR equation for the integrated gluon distribution. The model
based on nonlinear DGLAP (NLDGLAP) shows a correspondence with the
BK equation solution that calculates the values of the integrated
gluon density [21-27] at the initial point.\\
To determine the appropriate behavior we need to specify the
shadowing corrections to the gluon distribution at the initial
point. The unshadowed distribution to the gluon ($g^{u}$) at
$Q_{0}^{2}$ has to modifying for $x<x_{0}$ as
\begin{eqnarray}
xg(x,Q^{2}_{0})=xg^{u}(x,Q^{2}_{0})[1+\theta(x_{0}-x)[xg^{u}(x,Q^{2}_{0})\nonumber\\
-xg^{u}(x_{0},Q^{2}_{0})]/xg_{sat}(x,Q^{2}_{0})]^{-1},
\end{eqnarray}
where
$xg_{sat}(x,Q^{2})=\frac{16R^{2}Q^{2}}{27\pi\alpha_{s}(Q^{2})}$ is
the value of the gluon which would saturate the unitarity limit in
the leading shadowing approximation. The value of $R$ is depends
on how the gluon ladders couple to the proton, or on how the
gluons are distributed within the proton. $R$ will be of the order
of the proton radius $(R\simeq5\hspace{0.1cm} GeV^{-1})$ if the
gluons are spread throughout the entire nucleon, or much smaller
$(R\simeq2\hspace{0.1cm} GeV^{-1})$ if gluons are concentrated in
hot- spot [28] within the proton. Equation (2) satisfies the
requirements expected for the shadowing corrections to the gluon
distribution function in $Q_{0}^{2}$. It reduces to the unshadowed
form $xg^{u}$ when shadowing is negligible; that is, when
$xg_{sat}{\rightarrow}\infty$. In this region, the interaction of
gluons is negligible and we may continue to use evolution
equations linear in $g(x,Q^{2})$. However, at sufficiently
small-$x$, two gluons in different cascades may interact,
generally fusion the gluon ladders together and so decrease the
gluon density. Therefore shadowing contributions can no longer
be neglected.\\

This paper is organized as follows. In Sect.2, we derive the LO
master formula to extract the gluon distribution at small-$x$ from
the initial condition.  Section 3 is devoted to the analytical
solution of the master equation for the gluon distribution
functions using a Laplace-transform technique. In section 4 we
consider the nonlinear corrections to the gluon distribution
function with respect to the GLR-MQ equation using a Laplace
method. In the last section, we calculate numerically the
 gluon distribution functions from the standard DGLAP and NLDGLAP evolution equation  with respect to the  GLR-MQ corrections, respectively. Our
conclusion is
given in Sect.6.\\
\subsection{2. LO master formula}
The LO DGLAP  equation for the evolution of the gluon distribution
function at small-$x$ can be written as
\begin{eqnarray}
\frac{{\partial}G(x,Q^{2})}{{\partial}{\ln}Q^{2}}|_{DGLAP}=\nonumber\\
\frac{\alpha_{s}}{4\pi}\{G(x,Q^{2})(\frac{1}{3}(33-2n_{f})+12{\ln}(1-x))\nonumber\\
+12x[\int_{x}^{1}G(z,Q^{2})(\frac{z}{x}-2+\frac{x}{z}-(\frac{x}{z})^{2})\frac{dz}{z^{2}}\nonumber\\
+\int_{x}^{1}(G(z,Q^{2})-G(x,Q^{2}))\frac{z}{z-x}\frac{dz}{z^{2}}]\},
\end{eqnarray}
where $n_{f}$ being the number of active quark flavors (and
$G(x,Q^{2})=xg(x,Q^{2})$). In Eq.(3), the LO $g{\rightarrow}g$
splitting functions after the convolution integral which contains
plus prescription, $()_{+}$, are given by
\begin{eqnarray}
K_{1}(x)=\frac{1}{x}-2+x-x^{2},
\end{eqnarray}
and
\begin{eqnarray}
K_{2}(x)=\frac{1}{1-x}.
\end{eqnarray}
 Note
that Eq.(3) continue to hold at next-to-leading order with
modified gluon splitting function. To extract the shadowing gluon
distribution function, we first should be able to derive the DGLAP
equation for the gluon
distribution function using a Laplace-transform technique. This issue is the subject of the next section.\\
\subsection{3. Solution of the DGLAP equation for $G(x,Q^{2})$}
To solve  the standard DGLAP evolution equation for the gluon
distribution function  at small-$x$, we follow the procedure that
was used by BDM [13] and employ the Laplace-transform method to
solve this equation. Now we use the coordinate transformation
\begin{eqnarray}
\upsilon &{\equiv}& \ln(1/x).
\end{eqnarray}
Then the functions $\hat{G}$, $\hat{K}$, and $\hat{\mathcal{G}}$
in $\upsilon$-space are define by
\begin{eqnarray}
\hat{G}(\upsilon,Q^{2}) &{\equiv}& \hat{G}(e^{-\upsilon},Q^{2})
\end{eqnarray}
\begin{eqnarray}
\hat{K}(\upsilon) &{\equiv}& \hat{K}(e^{-\upsilon})
\end{eqnarray}
\begin{eqnarray}
\hat{\mathcal{G}}(\upsilon,Q^{2}) &{\equiv}&
\frac{\alpha_{s}}{4\pi}\frac{{\partial}\hat{G}(e^{-\upsilon},Q^{2})}{{\partial}{\ln}Q^{2}}.
\end{eqnarray}
Explicitly from Eqs.(3-5), we have
\begin{eqnarray}
\hat{k}(\upsilon)=\beta_{0}+12\ln(1-e^{-\upsilon}),
\end{eqnarray}
\begin{eqnarray}
\hat{K_{1}}(\upsilon) &=&
e^{\upsilon}-2+e^{-\upsilon}-e^{-2\upsilon}
\end{eqnarray}
and
\begin{eqnarray}
\hat{K_{2}}(\upsilon) &=&\frac{1}{1-e^{-\upsilon}}.
\end{eqnarray}
 In this representation, Eq.(3) reduces to this form
\begin{eqnarray}
\hat{\mathcal{G}}(\upsilon,Q^{2})=\hat{G}(\upsilon,Q^{2})\hat{k}(\upsilon)\nonumber\\
+12[\int_{0}^{\upsilon}\hat{G}(\omega,Q^{2})e^{-(\upsilon-\omega)}\hat{K}_{1}(\upsilon-\omega) \nonumber\\
+\int_{0}^{\upsilon}(\hat{G}(\omega,Q^{2})-\hat{G}(\upsilon,Q^{2}))e^{-(\upsilon-\omega)}\hat{K}_{2}(\upsilon-\omega)],
\end{eqnarray}
or
\begin{eqnarray}
\hat{\mathcal{G}}(\upsilon,Q^{2})&=&\hat{G}(\upsilon,Q^{2})\hat{k}(\upsilon)
+12[\int_{0}^{\upsilon}\hat{G}(\omega,Q^{2})\hat{H}_{1}(\upsilon-\omega) \nonumber\\
&&+\int_{0}^{\upsilon}(\hat{G}(\omega,Q^{2})-\hat{G}(\upsilon,Q^{2}))\hat{H}_{2}(\upsilon-\omega)]\nonumber\\
\end{eqnarray}
where
\begin{eqnarray}
\hat{H}_{1}(\upsilon)&{\equiv}&
e^{-\upsilon}\hat{K}_{1}(\upsilon)=1-2e^{-\upsilon}+e^{-2\upsilon}-e^{-3\upsilon}
\end{eqnarray}
and
\begin{eqnarray}
\hat{H}_{2}(\upsilon)&{\equiv}&
e^{-\upsilon}\hat{K}_{2}(\upsilon)=\frac{e^{-\upsilon}}{1-e^{-\upsilon}}.
\end{eqnarray}
We introduce the notation that the Laplace transformation of a
function $\hat{H}(\upsilon)$ is given by $h(s)$, where
\begin{eqnarray}
h(s)&{\equiv}&
{\mathcal{L}}[\hat{H}(\upsilon);s]=\int_{0}^{\infty}\hat{H}(\upsilon)e^{-s\upsilon}d\upsilon.
\end{eqnarray}
Then  Eq.(14) reduced to the form
\begin{eqnarray}
{\mathcal{G}}(s,Q^{2})={\mathcal{L}}[\hat{G}(\upsilon,Q^{2})\hat{k}(\upsilon);s]
+12{G}(s,Q^{2}){h}_{1}(s)\nonumber\\
+12{\mathcal{L}}[\int_{0}^{\upsilon}(\hat{G}(\omega,Q^{2})-\hat{G}(\upsilon,Q^{2}))\hat{H}_{2}(\upsilon-\omega);s],
\end{eqnarray}
where we used the following property for Laplace transformation
\begin{eqnarray}
{\mathcal{L}}[\int_{0}^{\upsilon}\hat{G}(\omega,Q^{2})\hat{H}(\upsilon-\omega);s]
={G}(s,Q^{2}){h}(s).
\end{eqnarray}
Let us note that some sentences are very small at the limit low
Bjorken scaling values, therefore we can estimate $G(s,Q^{2})$ as
follow
\begin{eqnarray}
\frac{{\partial}G(s,Q^{2})}{{\partial}{\ln}Q^{2}}{\simeq}
\frac{\alpha_{s}}{4\pi}(\beta_{0}+12{h}_{1}(s)){G}(s,Q^{2}).
\end{eqnarray}
Integrating both part of this equation, then Eq.(20) reduced to
this form
\begin{eqnarray}
G(s,Q^{2})=G(s,Q_{0}^{2})\exp[(1+\frac{12}{\beta_{0}}{h}_{1}(s))\eta(Q^{2})]
\end{eqnarray}
where
$\eta(Q^{2})=\int_{Q_{0}^{2}}^{Q^{2}}\frac{1}{{\ln}\frac{Q^{2}}{\Lambda^{2}}}d
{\ln}Q^{2}$.\\
In order to find solution for $G(x,Q^{2})$, using the inverse
Laplace transformation
${\mathcal{L}}^{-1}[G(s,Q^{2});\upsilon]=\hat{G}(\upsilon,Q^{2})$.
Then we find that
\begin{eqnarray}
{\mathcal{L}}^{-1}[G(s,Q^{2});\upsilon]={\mathcal{L}}^{-1}[G(s,Q_{0}^{2})e^{[(1+\frac{12}{\beta_{0}}{h}_{1}(s))\eta(Q^{2})]};\upsilon]\nonumber\\
\end{eqnarray}
and finally
\begin{eqnarray}
\hat{G}(\upsilon,Q^{2})=e^{\eta(Q^{2})}\int_{0}^{\upsilon}\hat{G}(\omega,Q_{0}^{2})\hat{J}(\upsilon-\omega)d\omega.
\end{eqnarray}
where $\hat{J}(\upsilon)$ is defined by
\begin{eqnarray}
\hat{J}(\upsilon)&{\equiv}&{\mathcal{L}}^{-1}[e^{\zeta{h}_{1}(s)};\upsilon].
\end{eqnarray}
and $ {\zeta} {\equiv} \frac{12\eta(Q^{2})}{\beta_{0}}$.\\
Eq.(23) can be regarded as the equation in terms of $s$ for
$e^{\zeta{h}_{1}(s)}$. In order to solve this equation we take the
first term of ${h}_{1}(s)$, i.e. $\hat{J}(\upsilon)$ obtained in
terms of the Dirac delta and Bessel functions of the first kind,
as
\begin{eqnarray}
\hat{J}(\upsilon)=\delta(\upsilon)
+\frac{\sqrt{\zeta}}{\sqrt{\upsilon}}BesselI(1,2\sqrt{\zeta}\sqrt{\upsilon}),\nonumber\\
\end{eqnarray}
and for the other terms we have
\begin{eqnarray}
{\mathcal{L}}^{-1}[\exp(\frac{-2\zeta}{s+1});\upsilon]=e^{-\upsilon}(\delta(\upsilon)\nonumber\\
+\frac{\sqrt{-2\zeta}}{\sqrt{\upsilon}}BesselI(1,2\sqrt{-2\zeta}\sqrt{\upsilon})),
\end{eqnarray}
\begin{eqnarray}
{\mathcal{L}}^{-1}[\exp(\frac{\zeta}{s+2});\upsilon]=e^{-2\upsilon}(\delta(\upsilon)\nonumber\\
+\frac{\sqrt{\zeta}}{\sqrt{\upsilon}}BesselI(1,2\sqrt{\zeta}\sqrt{\upsilon})),
\end{eqnarray}
\begin{eqnarray}
{\mathcal{L}}^{-1}[\exp(\frac{-\zeta}{s+3});\upsilon]=e^{-3\upsilon}(\delta(\upsilon)\nonumber\\
+\frac{\sqrt{-\zeta}}{\sqrt{\upsilon}}BesselI(1,2\sqrt{-\zeta}\sqrt{\upsilon})).
\end{eqnarray}
Combining (25) with (23) and using the properties of Dirac delta
function, we therefore obtain an explicit solution for the gluon
distribution function in $x$ space in terms of the initial
condition
\begin{eqnarray}
G(x,Q^{2})=e^{\eta(Q^{2})}[G(x,Q_{0}^{2})+\int_{x}^{1}G(z,Q_{0}^{2}){\times}\nonumber\\
\frac{\sqrt{\zeta}}{\sqrt{{\ln}\frac{z}{x}}}BesselI(1,2\sqrt{\zeta}\sqrt{{\ln}\frac{z}{x}})\frac{dz}{z}].
\end{eqnarray}
It should be noted that the size of Eqs.(26-28) are very small at
low-$x$ values. Eq.(29) means that we can use the Laplace
transform method for the solution of the DGLAP evolution equation
at small-$x$ values as the gluon distribution is dominant at this
region. This result is completely general and gives the evolution
solution for the gluon distribution function once the initial
gluon distribution function is known. In the next section we
propose this method for determining of  the shadowing corrections
to the gluon distribution function with respect to the GLR-MQ
 equation.\\

\subsection{4. Approximate solution of the GLR-MQ  equation}

The GLR-MQ equation slow down the $Q^{2}$ evolution of gluons from
the standard DGLAP behavior. Theses corrections arise from fusion
of two gluon ladders, and they modify the evolution equations of
gluons as  gluons have a high density at small-$x$ according to
the equation (1). The main purpose of this paper is to find an
approximate solution for the non-linear DGLAP evolution equation
in the saturation effects region. It is useful to consider the
GLR-MQ  equation in a $\upsilon$ space. In $\upsilon$ space, we
have defined the Laplace transform of the
$\int\frac{dy}{y}[G(y,Q^{2})]^{2}{\equiv}{\int}dw[\widehat{G}(w,Q^{2})]^{2}=\frac{1}{s}\mathcal{L}[(\widehat{G}(\upsilon,Q^{2}))^{2};s]$
to be less than $\frac{1}{s}[G(s,Q^{2})]^{2}$. Therefore in this
limit, we take the Laplace transform of (1), by follow
\begin{eqnarray}
\frac{{\partial}G(s,Q^{2})}{{\partial}{\ln}Q^{2}}&{\simeq}&\frac{{\partial}G(s,Q^{2})}{{\partial}{\ln}Q^{2}}|_{DGLAP}\nonumber\\
&&-\frac{81\alpha^{2}_{s}}{16R^{2}Q^{2}}\frac{1}{s}[G(s,Q^{2})]^{2},
\end{eqnarray}
 According to Eq.(20), we
rewrite Eq.(30) in terms of the  gluon transformed in $s$-space
\begin{eqnarray}
\frac{{\partial}G(s,Q^{2})}{{\partial}{\ln}Q^{2}}&=&{\psi}G(s,Q^{2})-{\varphi}G^{2}(s,Q^{2}),
\end{eqnarray}
where $\psi =
(1+\frac{12h_{1}(s)}{\beta_{0}})\frac{1}{{\ln}\frac{Q^{2}}{\Lambda^{2}}}$
and $\varphi =
\frac{81\pi^{2}}{s\beta^{2}_{0}R^{2}Q^{2}(\ln\frac{Q^{2}}{\Lambda^{2}})^{2}}$.
One may write the solution in fact directly in terms of the
initial condition, as
\begin{eqnarray}
G(s,Q^{2})&=&G(s,Q_{0}^{2})e^{\int_{Q_{0}^{2}}^{Q^{2}}{\psi}d{\ln}Q^{2}}\\
&&{\times}[1+G(s,Q_{0}^{2})\int_{Q_{0}^{2}}^{Q^{2}}e^{\int{\psi}d{\ln}Q^{2}}{\varphi}d{\ln}Q^{2}]^{-1}.\nonumber
\end{eqnarray}
Further, we can rewrite Eq.(32) by
\begin{eqnarray}
G(s,Q^{2})&=&G(s,Q_{0}^{2})e^{\int_{Q_{0}^{2}}^{Q^{2}}{\psi}d{\ln}Q^{2}}-G^{2}(s,Q_{0}^{2})e^{\int_{Q_{0}^{2}}^{Q^{2}}{\psi}d{\ln}Q^{2}}\nonumber\\
&&{\times}\int_{Q_{0}^{2}}^{Q^{2}}e^{\int{\psi}d{\ln}Q^{2}}{\varphi}d{\ln}Q^{2}+{\mathcal{O}}(2)-{\mathcal{O}}(3)+...\nonumber\\
\end{eqnarray}
We will generally be able to calculate the inverse Laplace
transform of $G(s,Q^{2})$ explicitly and transforming back into
$x$ space. Finally the shadowing corrections to the gluon
distribution function $G(x,Q^{2})$ obtained in terms of the
initial function $G(x,Q_{0}^{2})$, assumed to be known, by
\begin{eqnarray}
G(x,Q^{2})&=&[Eq.29]-\int_{\chi}^{1}G^{2}(z,Q_{0}^{2})Fe^{Y}\nonumber\\
&&{\times}BesselI(0,2\sqrt{U}\sqrt{{\ln}\frac{z}{x}})\frac{dz}{z}\nonumber\\
&&+{\mathcal{O}}(2)-{\mathcal{O}}(3)+...
\end{eqnarray}
In the above equation, the Bessel functions obtained from the
property for inverse Laplace transformation
\begin{eqnarray}
{\mathcal{L}}^{-1}[\frac{F}{s}e^{Y}e^{\frac{U}{s}};\upsilon]=Fe^{Y}
BesselI(0,2\sqrt{U}\sqrt{\upsilon})),
\end{eqnarray}
where the functions $F, U$ and $Y$ obtained from the below
equation after integration and some rearranging
\begin{eqnarray}
Fe^{Y}e^{\frac{U}{s}}&{\equiv}&
e^{\int^{Q^{2}}_{Q_{0}^{2}}{\psi}d{\ln}Q^{2}}
\int^{Q^{2}}_{Q_{0}^{2}}e^{(1+\frac{12h_{1}(s)}{\beta_{0}}){\ln}{\ln}\frac{Q^{2}}{\Lambda^{2}}}
{\varphi}d{\ln}Q^{2}.\nonumber\\
\end{eqnarray}
This method can be used for order ${\mathcal{O}}(2)$ and other
orders, as they are very small at low-$x$. This solution in the
form of Eq.(34) has a very clear physical meaning. We expect that
gluon correlations in the parton cascade at the region of the
small-$x$ tamed the behavior of the gluon distribution function.
Therefore we observed  that the
terms add to Eq.(29) on r.h.s of Eq.(34) are due to the saturation effects.\\
\subsection{5.Numerical Estimates}
In this paper, we found two analytical solutions for the standard
DGLAP  and NLDGLAP evolution equations for the gluon distribution
function inside the proton. These equations (Eqs.29 and 34) are
directly related to the initial conditions and they are
independent of the gluon behavior at small-$x$. We determined the
unshadowing and shadowing corrections to the gluon distribution
functions using a Laplace-transform method in Eqs.29 and 34,
respectively. Despite the analytical calculations that led us to
the solutions in the forms of Eqs.(29) and (34), we will solved
these equations numerically. In these calculations, we use the LO
form of  $\alpha_{s}(Q^{2})$ which is defined by
\begin{eqnarray}
\alpha_{s}(Q^{2})&{\equiv}&\frac{12\pi}{25\ln(\frac{Q^{2}}{\Lambda^{2}})},
\end{eqnarray}
for four active flavors, where $\Lambda$ is the QCD cut- off
parameter. The input gluon parameterization can be taken  from the
BDM model [12] at the starting scale $Q_{0}^{2}=5~
\mathrm{GeV}^{2}$. In Fig.1, we show shadowing corrections to the
gluon distribution function  determined from Eq.(2) as a function
of $x$ at the initial scale $Q_{0}^{2}$ based on the size of the
target which the gluons populate. We observed that, as $x$
decreases, the singularity behavior of the gluon functions are
tamed by shadowing effects. The top curves (solid and dot) show
the small-$x$ behavior of the gluon distribution when shadowing is
neglected, and the lower curves (dash) show the effect of the
shadowing contribution for $R=5$ and $4~ GeV^{-1}$. We compared
our results with BK [15-17](dash-dot) curve and GJR
parameterization [29] (solid).\\
In Fig.2, we show the unshadowed corrections to the  gluon
distribution function $G(x,Q^{2})$ determined from Eq.(29) as a
function of $x$ for two different values of $Q^{2}$, namely
$Q^{2}=20 $ and $Q^{2}=50~ GeV^{2}$ with respect to the
Laplace-transform method. We compared our results with GJR
parameterization [29] and BDM model [12]. By using this method, we
show that our results are general and independent of the gluon
behavior at small-$x$. As can be seen, the values of the gluon
distribution increase as $x$ decreases.\\
In Fig.3, we show the shadowing corrections to the gluon
distribution function determined  from Eq.(34). To evaluate the
NLDGLAP equation, by using a Laplace-transform method, we have
generalized the BDM [12-13] approach from linear to nonlinear case
at small-$x$. The results are shown for $Q^{2}=20 $ and $Q^{2}=50~
GeV^{2}$.  The top curve (squared) shows the small-$x$ behavior of
the gluon distribution when shadowing is neglected, and the lower
curve (circled) shows the effects of the shadowing corrections  at
the extreme "hot- spot" point [30] with
$R=2\hspace{0.1cm}GeV^{-1}$. Our results compared with the
integrated gluon density function, which is denote as BK curve.
These results show that gluon density behavior increase as x
decreases, this corresponds with PQCD fits at low-$x$, but this
behavior is tamed with respect to nonlinear terms at the GLR-MQ
equation. This figure compares the results of two calculations,
based on the linear and nonlinear equations. The differences are
not large, however there is some suppression due to the
nonlinearity at smallest values of $x{\leq}10^{-4}$. It is clear
that a new evolution equation for the gluon density relevant for
the region of small-$x$. It show that screening effects
 are provided by multiple gluon interaction which leads to the nonlinear terms in the DGLAP equation.\\
In order to compare our results with the experimental data, we
follow  the role of shadowing correction to the evolution of the
singlet quark distribution with respect to GLR-MQ equations.
Indeed, the DGLAP evolution equation of the $Q^{2}$ logarithmic
slope of $F_{2}$ at low $x$, at leading order, is directly
proportional to the gluon structure function. But, according to
these corrections (fusion of two gluon ladders), the shadowing
corrections to the evolution of the structure function with
respect to ${\ln}Q^{2}$ corresponds with a modified DGLAP
evolution equation [10-11,30-32]. At low $x$, the nonsinglet
contribution is negligible and can be ignord, so that
\begin{eqnarray}
\frac{{\partial}F_{2}(x,Q^{2})}{{\partial}{\ln}Q^{2}}&=&\frac{{\partial}F_{2}(x,Q^{2})}{{\partial}{\ln}Q^{2}}|_{DGLAP}\nonumber\\
&&-\frac{5}{18}\frac{27\alpha_{s}^{2}}{160R^{2}Q^{2}}
[xg(x,Q^{2})]^{2},
\end{eqnarray}
where the first term is the standard DGLAP evolution equation.\\
To find an analytical solution for the shadowing corrections to
the derivative of $F_{2}$ at low $x$, we intend to use the gluon
functions obtained in Eq.38. We show a plot of
$\frac{{\partial}F_{2}(x,Q^{2})}{{\partial}{\ln}Q^{2}}$ in Fig.4
for a the values of $Q^{2}=20 GeV^{2}$ and $50 GeV^{2}$ at the
points $R=2 GeV^{-1}$ and $4 GeV^{-1}$, compared to the values
measured by the H1 collaboration
 [33-34] and a QCD fit based on ZEUS data [35]. From this figure one
 can see the GLR-MQ equation is tamed behavior with respect to gluon
 saturation as $Q^{2}$ increases. Consequently, derivative of the
 shadowing corrections to the structure function with respect to ${\ln}Q^{2}$ at fixed
 $x$ is close to H1 data and to the QCD analysis fit. This shadowing
 correction suppresses the rate of growth in comparison with the
 DGLAP approach.\\

\subsection{6.Conclusions}
In this paper, we studied the effects of adding the nonlinear
GLR-MQ corrections to the LO DGLAP evolution equation, by using a
Laplace-transform method suggested by the BDM [12-14], in order to
determine the small-$x$ behavior of the gluon distribution
$xg(x,Q^{2})$ of the proton. In this way we were able to study the
interplay of the singularity behavior, generated by the linear
term of the equation, with the taming of this behavior by the
nonlinear shadowing term. The nonlinear effects to the gluon
distribution are found to play an increasingly important at
$x{\leq}10^{-4}$. The nonlinearities, however, vanish rapidly at
large values of $x$. We observed that, as $x$ decreases, the
singularity behavior of the gluon function is tamed by shadowing
effects.  Our results compared with the gluon distribution
obtained in a QCD analysis and with the BK equation. The obtained
results  are comparable with the results obtained by basing on the
BK equation. Based on our present calculations we conclude that
the behavior of
$\frac{{\partial}F_{2}(x,Q^{2})}{{\partial}{\ln}Q^{2}}$ as
measured at HERA, tamed based on gluon saturation at low $x$. The
expansion of the results from LO to NLO approximation can be
done as a new research job in future.\\

\section{References}

1.Yu.L.Dokshitzer, Sov.Phys.JETP {\textbf{46}}, 641(1977).\\
2.G.Altarelli and G.Parisi, Nucl.Phys.B \textbf{126}, 298(1977).\\
3.V.N.Gribov and L.N.Lipatov, Sov.J.Nucl.Phys. \textbf{15},
438(1972).\\
4. A.De Rujula \textit{et al}., phys.Rev.D \textbf{10},
1649(1974).\\
5. R.D.Ball and S.Forte, Phys.Lett.B \textbf{335}, 77(1994).\\
6. A.H.Mueller and J.Qiu, Nucl.Phys.B\textbf{268}, 427(1986).\\
 7. J.Kwiecinski and A.M.Stasto, Phys.Rev.D\textbf{66},
 014013(2002).\\
8. L.V.Gribov, E.M.Levin and M.G.Ryskin, Phys.Rep.\textbf{100},
 1(1983).\\
 9. A.L.A.Filho. M.B.Gay Ducati and V.P.Goncalves,
 Phys.Rev.D\textbf{59},054010(1999).\\
10. E.Laenen and E.Levin,Nucl.Phs.B.\textbf{451},207(1995).\\
11. K.J.Eskola, et.al., Nucl.Phys.B\textbf{660}, 211(2003).\\
12.M.M.Block, L.Durand, D.W.McKay, Phys.Rev.D\textbf{77},
094003(2008).\\
13.M.M.Block, L.Durand, D.W.McKay, Phys.Rev.D\textbf{79},
014031(2009).\\
14.M.M.Block, Eur.Phys.J.C\textbf{65},
1(2010).\\
15. I.I.Balitsky, Nucl.Phys.B\textbf{463}, 99(1996).\\
16.Yu.V.Kovchegov, Phys.Rev.D\textbf{60}, 034008(1999).\\
17.Yu.V.Kovchegov, Phys.Rev.D\textbf{61}, 074018(2000).\\
18. E.A.Kuraev, L.N.Lipatov and V.S.Fadin, Sov.Phys.JETP
{\bf{44}}, 443(1976).\\
19. E.A.Kuraev, L.N.Lipatov and V.S.Fadin, Sov.Phys.JETP
{\textbf{45}}, 199(1977).\\
20. Y.Y.Balitsky and L.N.Lipatov, Sov.Journ.Nucl.Phys. {\textbf{28}}, 822(1978).\\
21. K.Golec-Biernat,
L.Motyka, and A.M.Stasto, Phys.Rev.D\textbf{65}, 074037(2002).\\
22. J. Kwiecinski, A.D. Martin, P.J. Sutton, Phys. Rev. D
\textbf{44}, 2640 (1991).\\
23. A.J. Askew, J. Kwiecinski, A.D. Martin, P.J. Sutton,
Phys.Rev. D \textbf{47}, 3775 (1993).\\
24. S.Bondarenko, Phys.Lett.B\textbf{665}, 72(2008).\\
25. K.Kutak and A.M.Stasto, Eur.Phys.J.C\textbf{41}, 343(2005).\\
26. M.Kazlov and E.Levin, Nucl.Phys. A\textbf{764}, 498 (2006).\\
27. M.A.Kimber,
J.Kwiecinski and A.D.Martin,  Phys.Lett. B\textbf{508}, 58(2001).\\
28.E.M.Levin and M.G.Ryskin, Phys.Rep.\textbf{189}, 267(1990).\\
29.M. Gluck, P. Jimenez-Delgado, E. Reya, Eur.Phys.J.C {\bf53},355
({\bf2008}).\\
30. A.H. Mueller,arXiv:hep-ph/0111244(2001).\\
31. E.Gotsman,et.al., Nucl.Phys.B\textbf{539}, 535(1999).\\
32. E.Gotsman,et.al., Phys.Lett.B\textbf{500}, 87(2001).\\
33. $H1$ Collab., C.Adloff \textit{et al}., Eur.Phys.J.C
\textbf{13}, 609(2000).
34. $H1$ Collab., C.Adloff \textit{et al}., Eur.Phys.J.C \textbf{21}, 33(2001).\\
35. E.L.Berger, M.M.Block and C.I Tan, Phys.Rev.Lett\textbf{98},
242001(2007).\\

\begin{figure}
\includegraphics[width=0.5\textwidth]{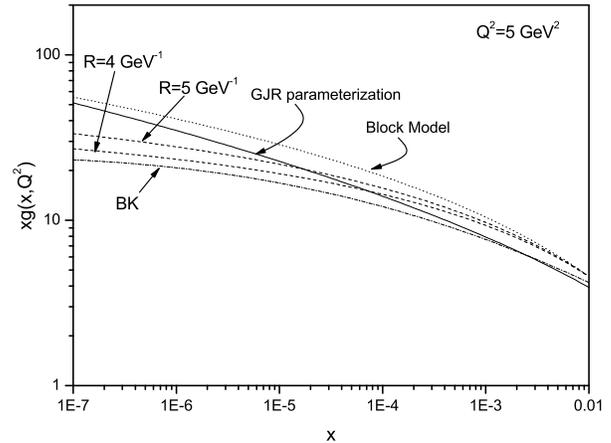}
\caption{ The shadowing corrections to the gluon distribution
function at $Q_{0}^{2}=5\hspace{0.1cm} GeV^{2}$ from the solution
of the
 initial condition shadowing effects based on Eq.2 with the boundary condition at $x_{0}=10^{-2}$.
  The  dot curve is the BDM model [12]
  with unshadowed corrections
   and dash curves are our results with the shadowing corrections  included with $R=4\hspace{0.1cm}GeV^{-1}$ and
   $R=5\hspace{0.1cm}GeV^{-1}$  that compared with BK [15-17,24-27] integrated gluon density function
   (dash-dot) and GJR [29] parameterization (solid) at small-$x$.}\label{Fig1}
\end{figure}
\begin{figure}
\includegraphics[width=0.5\textwidth]{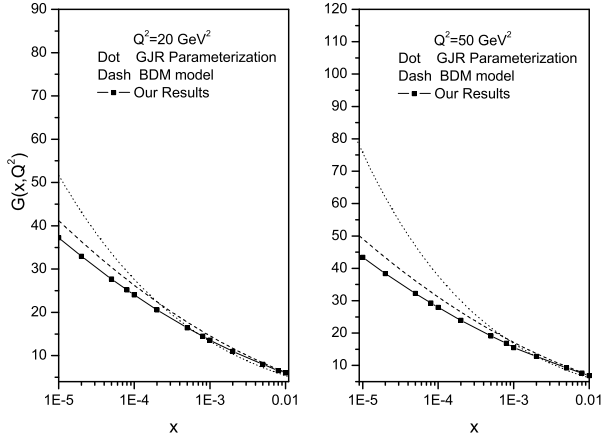}
\caption{ Unshadowed LO gluon distribution function at $Q^{2}=20$
and $50 ~ GeV^{2}$ according to the gluon distribution input [12]
compared with results BDM model [12] and GJR [29] parameterization
at small-$x$. }\label{Fig2}
\end{figure}
\begin{figure}
\includegraphics[width=0.5\textwidth]{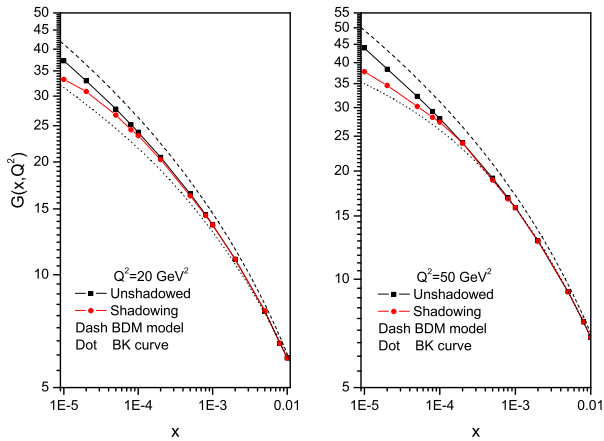}
\caption{Shadowing corrections to the  gluon distribution function
 included with $R=2\hspace{0.1cm}GeV^{-1}$ at  $Q^{2}=20$ and $50 ~ GeV^{2}$ compared with results BDM model [12] and BK
[15-17,24-27] equation at small-$x$ . }\label{Fig3}
\end{figure}
\begin{figure}
\includegraphics[width=0.5\textwidth]{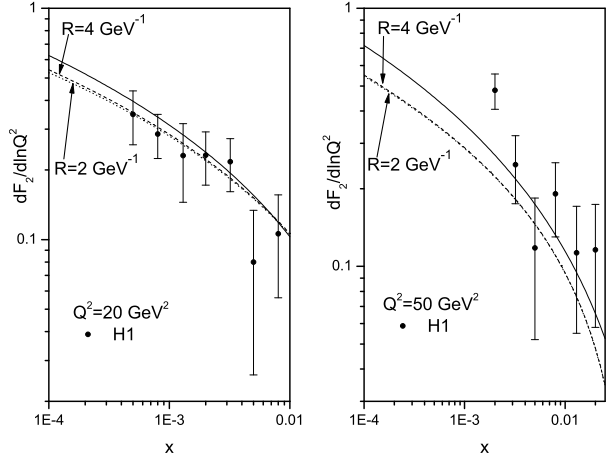}
\caption{A plot of the derivative of the shadowing corrections to
the structure function with respect to ${\ln}Q^{2}$ for $Q^{2}=20
GeV^{2} $ and $50 GeV^{2}$ at the points $R=2 GeV^{-1}$
(Dot-line)and $4 GeV^{-1}$ (Dash-line), compared to data from H1
Collab.[33-34] (circles) with total error, and also a QCD fit to
ZEUS data [3](solid-line). }\label{Fig4}
\end{figure}


\end{document}